\documentclass[12pt]{article}
\usepackage{amsmath}
\usepackage{amsfonts}   
\usepackage{amssymb}
\def\a{\alpha}
\def\b{\beta}

\def\d{\delta}
\def\r{\rho}
\def\s{\sigma}
\def\m{\mu}
\def\n{\nu}
\def\p{\partial}
\begin{document}
\date{}
\title{\textbf{Diffeomorphism Symmetry in the Lagrangian Formulation of Gravity}}
\author{{Saurav Samanta}\thanks{E-mail: saurav@bose.res.in}\\
\\\textit{S.~N.~Bose National Centre for Basic Sciences,}
\\\textit{JD Block, Sector III, Salt Lake, Kolkata-700098, India}}
\maketitle
                                                                                
\begin{quotation}
\noindent \normalsize
Starting from a knowledge of certain identities in the Lagrangian description, the diffeomorphism transformations of metric and connection are obtained for both the second order (metric) and the first order (Palatini) formulations of gravity. The transformation laws of the connection and the metric are derived independently in the Palatini formulation in contrast to the metric formulation where the gauge variation of the connection is deduced from the gauge variation of the metric. 
\end{quotation}
\section{Introduction}
Classical description of gravity is best formulated by general relativity theory where gravity is treated as curvature of space time instead of an external force. In order to build up the whole formulation one needs to introduce two logically independent concepts, one is metric and the other is connection. Metric is a symmetric second rank covariant tensor which is used to define an invariant length on the manifold and the connection is introduced to map the vectors of different tangent spaces, which in turn is used to define the covariant derivative. Though these two concepts are entirely independent, in the standard version of general relativity one demands two conditions i.e. torsion free nature of connection and the vanishing covariant derivative of metric (metric compatibility condition)\cite{carol}. The violation of the first condition together with a general asymmetric metric leads to many theoretical results in which Einstein himself was very much interested\cite{ein1,ein2}. The second condition, in a special formulation of general relativity known as Palatini formulation, is not assumed apriori but derived as an equation of motion\cite{pala}. Though this formulation gives the same result as the standard general relativity in vacuum, it leads to distinct results when spinors are coupled to gravity\cite{deser}.

A relativistic description of classical gravity whether it is standard (metric) formulation or Palatini formulation is essentially based on the physical concept of general invariance principle. It is also known that the general theory of relativity is invariant under local Poincare transformation. Thus one argues that localization of Poincare symmetry leads to the principle of equivalence which is the central concept of all gravitational theories. In this approach each space time point is associated with a set of local inertial coordinates, mutually related by Lorentz transformation, which enables one to formulate the Poincare gauge theory of gravity\cite{kibble,sci}.

But even without gauging the Poincare symmetry, the gauge (diffeomorphism) transformations of the fields can be obtained from the Einstein Hilbert action where the presence of gauge symmetry is indicated by the first class constraints of the theory. Following Dirac's algorithm\cite{dirac}, generator of the gauge or diffeomorphism transformations can be constructed from a linear combination of the constraints which finally gives the diffeomorphism transformation of the metric field\cite{cas,pradip}. An important tool in this formulation is the A-D-M (Arnowitt-Deser-Misner) decomposition\cite{adm} of space time where three space surfaces evolve along a time like direction. In fact it is well known that this decomposition plays a central role in all Hamiltonian formulations of general relativity.

A question therefore naturally arises whether it is possible to describe the diffeomorphism symmetries without recourse to the A-D-M decomposition. The point is that there is a Lagrangian version of discussing gauge symmetries using certain gauge identities\cite{gitman}. These gauge identities involve the Euler derivatives and the generators of the gauge transformations. This approach has been used in various contexts\cite{shirzad,rabin,rabina} including noncommutative gauge theories\cite{saurav}.

 In the present paper we show the existence of certain identities for both the metric and the Palatini formulations. From these identities the diffeomorphism transformations of the basic fields are calculated in a systematic manner.

This paper is organised as follows. In Section 2 we give a short discussion on the general method of analyzing the gauge symmetry in the Lagrangian frame work. Non-Abelian gauge theory, both in its second and first order versions are taken as examples to illustrate the method. The second order formulation of metric gravity is discussed in Section 3, whereas the first order Palatini formulation is analyzed in Section 4. Section 5 is for conclusions and after that a short appendix is added. 
\section{General Discussion}
It is a well known fact that a theory containing gauge symmetry possesses an identity involving the various Euler derivatives of the theory. This identity is called the gauge identity from which the gauge transformations of the fields can be calculated. We briefly discuss this method in the first part of this section for a general gauge theory and later we show how this method works for a non-Abelian gauge theory. The gauge transformations of the basic fields are calculated for both the second order and the first order (Palatini-like) forms of the theory.

To study the dynamics of a field from an action principle we consider a general Lagrangian involving only upto first order derivatives of the field of the form\footnote{We use the notation $x$ for the four vector $x^{\mu}=({\bf{x}},t)$.},
\begin{eqnarray}
S=\int\textrm{d}t \ L=\int \textrm{d}^4x \  \mathcal{L}
\left(q_{\alpha}({\bf{x}},t), \ \partial_iq_{\alpha}({\bf{x}},t), \ \partial_tq_{\alpha}({\bf{x}},t)\right)
\label{L}
\end{eqnarray}
where $\alpha$ denotes the number of fields. Also, it contains all other (e. g. symmetry) indexes relevant for the problem. An arbitrary variation of this action is written as
\begin{eqnarray}
\delta S=-\int \textrm{d}^4x \ \delta q^{\alpha}({\bf{x}},t)L_{\alpha}({\bf{x}},t).
\label{l}
\end{eqnarray}
The equations of motion are obtained by setting the Euler derivative $L$ to be zero,
\begin{eqnarray}
L_{\alpha}=0.
\end{eqnarray}
Now we vary the field $q^{\alpha}$ in the following way
\begin{eqnarray}
\delta q^{\alpha}({\bf{x}},t)=\sum_{s=0}^n(-1)^s\int\textrm{d}^3{\bf{z}} \ \frac{\partial^s\eta^b({\bf{z}},t)}{\partial t^s}\rho^{\alpha b}_{(s)}(x,z)
\label{a}
\end{eqnarray}
with $\eta$ and $\rho$ being the parameter and generator, respectively, of the transformation. Under this variation of the field, the variation of the action is written from (\ref{l}) as
\begin{eqnarray}
\delta S&=&-\int\textrm{d}^4x \ \int\textrm{d}^3{\bf{z}} \ \eta^b({\bf{z}},t)\rho^{\alpha b}_{(0)}(x,z)L_{\alpha}({\bf{x}},t)-\nonumber\\
&&\int \textrm{d}^4x \  \sum_{s=1}^n(-1)^s\int \textrm{d}^3{\bf{z}} \ \frac{\partial}{\partial t}\left(\frac{\partial^{s-1}\eta^b({\bf{z}},t)}{\partial t^{s-1}}\right)\rho^{\alpha b}_{(s)}(x,z)L_{\alpha}({\bf{x}},t)\nonumber\\
&=&-\int\textrm{d}^4x \ \int\textrm{d}^3{\bf{z}} \ \eta^b({\bf{z}},t)\rho^{\alpha b}_{(0)}(x,z)L_{\alpha}({\bf{x}},t)-\nonumber\\
&&\int \textrm{d}^4x \sum_{s=1}^n(-1)^{s-1}\int \textrm{d}^3{\bf{z}} \ \frac{\partial^{s-1}\eta^b({\bf{z}},t)}{\partial t^{s-1}}\frac{\partial}{\partial t}\left(\rho^{\alpha b}_{(s)}(x,z)L_{\alpha}({\bf{x}},t)\right)\nonumber\\
&=&-\int \textrm{d}^4z \ \eta^b({\bf{z}},t)\left(\int \textrm{d}^3{\bf{x}} \ \rho^{\alpha b}_{(0)}(x,z)L_{\alpha}({\bf{x}},t)\right)-\nonumber\\
&&\int \textrm{d}^4z \ \eta^b({\bf{z}},t)\left(\int \textrm{d}^3{\bf{x}} \ \frac{\partial}{\partial t}(\rho^{\alpha b}_{(1)}(x,z)L_{\alpha}({\bf{x}},t))\right)-\cdot\cdot\cdot
\label{mann}
\end{eqnarray}
We define a quantity\cite{gitman,shirzad}
\begin{eqnarray}
\Lambda^a({\bf{z}},t)=\left[\sum_{s=0}^n\int \textrm{d}^3{\bf{x}} \ \frac{\partial^s}{\partial t^s}\left(\rho^{\alpha a}_{(s)}(x,z)L_{\alpha}({\bf{x}},t)\right)\right].
\label{lam}
\end{eqnarray}
to write eq. (\ref{mann}) in the form
\begin{eqnarray}
\delta S=-\int\textrm{d}^4z \ \eta^a({\bf{z}},t)\Lambda^a({\bf{z}},t).
\label{bak}
\end{eqnarray} 
If the action does not change ($\delta S=0$) under the field transformation (\ref{a}) then it implies,
\begin{eqnarray}
\Lambda^a({\bf{z}},t)=0.
\end{eqnarray}

The last equality, which is called the gauge identity, must be true without use of any equation of motion. The gauge transformations of the fields are defined by eq. (\ref{a}) with $\rho$ being the generator. 

To illustrate the use of this general formalism in a particular example, we consider the non-Abelian gauge theory to obtain the gauge transformations of the fields. We take the standard second order action,
\begin{eqnarray}
S=\int \textrm{d}^4x \ [-\frac{1}{2}{\textrm{Tr}}F_{\mu\nu}(x)F^{\mu\nu}(x)]
\label{lag}
\end{eqnarray}
where the definition of the field strength tensor
\begin{eqnarray}
&&F_{\mu\nu}\equiv\partial_{\mu}A_{\nu}-\partial_{\nu}A_{\mu}+ig[A_{\mu},A_{\nu}]
\label{f}
\end{eqnarray}
is given in terms of the gauge field
\begin{eqnarray}
A_{\m}=A_{\m}^aT^a.
\end{eqnarray}
Here $T^a$ are the generators of a gauge group satisfying{\footnote{We choose the representation where the structure constant $f^{abc}$ is antisymmetric in all indexes and ${\textrm{Tr}}(T^aT^b)=\frac{1}{2}\d^{ab}$.}}
\begin{eqnarray}
[T^a,T^b]=if^{abc}T^c.
\label{fabc}
\end{eqnarray}
The Euler derivatives are obtained by varying the action with respect to the $A_{\mu}$ field,
\begin{eqnarray}
\delta S=-\int \textrm{d}^4x \ \delta A_{\mu}^{a}L^{\mu a}
\label{tm}
\end{eqnarray}
The Euler derivatives $L^{\mu a}$ thus obtained are given by,
\begin{eqnarray}
L^{\mu a}=-\left(\mathcal{D}_{\sigma}F^{\sigma\mu}\right)^a
\end{eqnarray}
leading to the familiar equation of motion,
\begin{eqnarray}
L^{\mu a}=0.
\label{lmua}
\end{eqnarray}
Here the covariant derivative $\mathcal{D}$ is defined in the adjoint representation,
\begin{eqnarray}
&&\mathcal{D}_{\mu}\xi=\partial_{\mu}\xi+ig[A_{\mu},\xi];\\
&&(\mathcal{D}_{\mu}\xi)^a=\partial_{\mu}\xi^{a}-gf^{abc}A^b_{\mu}\xi^{c}.
\label{D}
\end{eqnarray}
In order to obtain the gauge transformations it is necessary to find the gauge identity. This is easily done by observing, from the definitions of $L^{\mu}$ and $\mathcal{D}$, the following relation
\begin{eqnarray}
\Lambda^a=-\left(\mathcal{D}^{\mu}L_{\mu}\right)^a=0.
\label{lambda}
\end{eqnarray}
This relation is true irrespective of the equation of motion of the fields and is referred as the gauge identity. Comparing this identity with the general form (\ref{lam}) we get the following generators\cite{rabin}
\begin{eqnarray}
\rho^{b0a}_{(0)}(x,z)&=&-gf^{abc}\delta^3({\bf{x}}-{\bf{z}})A_0^c(x)
\label{a1}\\
\rho^{b0a}_{(1)}(x,z)&=&-\delta^{ab}\delta^3({\bf{x}}-{\bf{z}}).
\label{a2}\\
\rho^{bia}_{(0)}(x,z)&=&-\delta^{ab}\partial^{i{\bf{z}}}\delta^3({\bf{x}}-{\bf{z}})-\nonumber\\
 &&gf^{abc}\delta^3({\bf{x}}-{\bf{z}})A^{ic}(x)
\label{37}
\end{eqnarray}
Using these expressions of the generators, we now obtain the gauge transformations of the $A^{\m}$ field from (\ref{a})
\begin{eqnarray}
\delta A^{\m a}=\partial^{\m}\eta^{a}-gf^{abc}A^{\m b}\eta^{c}=(\mathcal{D}^{\m}\eta)^a.
\label{r2}
\end{eqnarray}
The gauge transformation of the field strength tensor $F_{\mu\nu}$ is calculated from the definition (\ref{f}) and (\ref{r2}),
\begin{eqnarray}
\d F^{\m\n}=ig[F^{\m\n},\eta].
\end{eqnarray}
We now consider the first order action of the non-Abelian gauge theory
\begin{eqnarray}
S=\int \textrm{d}^4x \ \frac{1}{2}{\textrm {Tr}}F_{\mu\nu}F^{\mu\nu}-{\textrm {Tr}}F_{\mu\nu}(\p^{\m}A^{\n}-\p^{\n}A^{\m}+ig[A^{\m},A^{\n}])
\end{eqnarray}
where $A^{\m}$ and $F^{\mu\nu}$ are treated as independent fields. Following the same method, the Euler derivatives are calculated by varying the action with respect to $A^{\m}$ and $F^{\mu\nu}$
\begin{eqnarray}
\d S=-\int \textrm{d}^4x \ (\d A_{\m}L^{\m}+\d F_{\m\n}L^{\m\n}).
\end{eqnarray}
Invariance of the action ($\d S=0$) leads to the equations of motion
\begin{eqnarray}
&&L^{\m}\equiv\mathcal{D}_{\n}F^{\n\m}=0
\label{lmu}\\
&&L^{\m\n}\equiv\frac{1}{2}[F^{\m\n}-(\p^{\m}A^{\n}-\p^{\n}A^{\m}+ig[A^{\m},A^{\n}])]=0.
\label{lmunu}
\end{eqnarray}
Note that eq. (\ref{lmu}) which is the genuine equation of motion is identical with eq. (\ref{lmua}). The other equation (\ref{lmunu}) basically defines the field strength tensor.
A gauge identity is also found similar to the second order formulation
\begin{eqnarray}
\mathcal{D}_{\n}L^{\n}+ig[F^{\m\n},L_{\m\n}]=0
\label{991}
\end{eqnarray}
which can be verified using the definitions of $L^{\m}$ and $L^{\m\n}$. Expanding eq. (\ref{lam}) to explicitly write the two distinct Euler derivatives, we obtain the gauge identity as,
\begin{eqnarray}
\Lambda^a({\bf{z}},t)&=&\sum_s\int \textrm{d}^3{\bf{x}} \ \frac{\partial^s}{\partial t^s}\left(\rho^{b\mu a}_{(s)}(x,z)L^b_{\mu}({\bf{x}},t)\right)+\nonumber\\
&&\sum_s\int \textrm{d}^3{\bf{x}} \ \frac{\partial^s}{\partial t^s}\left(\rho^{b\mu\nu a}_{(s)}(x,z)L^b_{\mu\nu}({\bf{x}},t)\right).
\label{eq}
\end{eqnarray}
Comparing this with (\ref{991}) the generators are found out to be
\begin{eqnarray}
\rho^{b0a}_{(0)}(x,z)&=&-gf^{abc}\delta^3({\bf{x}}-{\bf{z}})A_0^c(x)
\label{b1}\\
\rho^{b0a}_{(1)}(x,z)&=&-\delta^{ab}\delta^3({\bf{x}}-{\bf{z}}).
\label{b2}\\
\rho^{bia}_{(0)}(x,z)&=&-\delta^{ab}\partial^{i{\bf{z}}}\delta^3({\bf{x}}-{\bf{z}})-\nonumber\\
 &&gf^{abc}\delta^3({\bf{x}}-{\bf{z}})A^{ic}(x)\label{38}\\
\rho^{b\m\n a}_{(0)}(x,z)&=&-gf^{abc}\delta^3({\bf{x}}-{\bf{z}})F^{\m\n c}(x).
\label{39}
\end{eqnarray}
The first three generators (\ref{b1},\ref{b2},\ref{39}) are same as the generators obtained in the second order formulation (\ref{a1},\ref{a2},\ref{37}). Eq. (\ref{39}) gives the only new generator.
The gauge transformations of the field $A_{\m}$ is now obtained from the relation (\ref{a})
\begin{eqnarray}
&&\delta A^{\m}=(\mathcal{D}^{\m}\eta).
\end{eqnarray}
The variation of the $F^{\m\n}$ is also calculated similarly
\begin{eqnarray}
\d F^{\m\n a}(x)&=&\int \textrm{d}^3z \ \eta^b(z)\r^{a\m\n b}_{(0)}(x,z)\\
&=&gf^{abc}\eta^bF^{\m\n c}.
\end{eqnarray}
Using (\ref{fabc}) we write the above equation as
\begin{eqnarray}
&&\d F^{\m\n}=ig[F^{\m\n},\eta].
\end{eqnarray}
In this way we obtain the gauge variation of the gauge field ($A^{\m}$) and the field strength ($F^{\m\n}$) in an independent manner for the first order formulation.

\section{Metric Formulation}
So far we were discussing the general method of analyzing the gauge symmetry in the Lagrangian framework and took non-Abelian gauge theory as our example to elaborate the whole procedure for both the second order and the first order descriptions. Now we are in a position to study the diffeomorphism symmetry of the general theory of relativity. In this section we study the second order formulation which is usually called the metric formulation. The less studied first order formulation i. e. the Palatini formulation, will be considered in the next section.

The Einstein-Hilbert action which describes the metric formulation of gravity is given by, 
\begin{eqnarray}
S&=&\int d^4x\mathcal{L}(g)\nonumber \\
&=&\int d^4x\sqrt{-g}R=\int d^4x\sqrt{-g}g^{\m\n}R_{\m\n}(g)
\label{cs}
\end{eqnarray}
where $R_{\m\n}$ is the Ricci tensor
\begin{eqnarray}
R_{\m\n}=\Gamma^{\lambda}_{\n\m,\lambda}-\Gamma^{\lambda}_{\lambda\m,\n}
+\Gamma^{\lambda}_{\n\m}\Gamma^{\s}_{\s\lambda}-\Gamma^{\s}_{\lambda\m}\Gamma^{\lambda}_{\n\s}.
\label{ricci}
\end{eqnarray}
The metric compatibility condition
\begin{eqnarray}
\nabla_{\r}g_{\m\n}=0
\label{metric}
\end{eqnarray}
 defines the Christoffel connection in terms of the metric components,
\begin{eqnarray}
\Gamma^{\r}_{\m\n}=\frac{1}{2}g^{\r\s}(g_{\n\s,\m}+g_{\m\s,\n}-g_{\m\n,\s}).
\label{Gamma}
\end{eqnarray}
Varying the action (\ref{cs}) with respect to the metric $g_{\m\n}$ we get the Euler derivative $L_{\m\n}$ i. e.
\begin{eqnarray}
\d S=\int L^{\m\n}\d g_{\m\n}
\end{eqnarray}
where the explicit form of  $L^{\m\n}$ is written as,
\begin{eqnarray}
L^{\m\n}=\sqrt{-g}G^{\m\n}=\sqrt{-g}(R^{\m\n}-\frac{1}{2}g^{\m\n}R)
\label{euler}
\end{eqnarray}
leading to the usual Einstein's equation, $L^{\m\n}=0$. Now to find the gauge identity we recall the Bianchi identity\cite{wineberg}
\begin{eqnarray}
\nabla_{\eta}R_{\lambda\m\n\kappa}+\nabla_{\n}R_{\lambda\m\kappa\eta}+\nabla_{\kappa}R_{\lambda\m\eta\n}=0
\label{bianchi}
\end{eqnarray}
which follows from the definition of the Riemann tensor
\begin{eqnarray}
R_{\lambda\m\n\kappa}=g_{\lambda\sigma}R^{\sigma}_{ \ \m\n\kappa}=g_{\lambda\sigma}(\Gamma^{\sigma}_{\m\kappa,\n}
-\Gamma^{\sigma}_{\m\n,\kappa}+\Gamma^{\eta}_{\m\kappa}\Gamma^{\sigma}_{\n\eta}-\Gamma^{\eta}_{\m\n}\Gamma^{\sigma}_{\kappa\eta}).
\end{eqnarray}
Contracting $\lambda$ with $\n$ and $\m$ with $\kappa$, in (\ref{bianchi}), using (\ref{metric}) we get
\begin{eqnarray}
\nabla_{\m}G^{\m\n}=0.
\end{eqnarray}
This contracted Bianchi which means that Einstein tensor $G^{\m\n}$ is divergence free is referred as the gauge identity in \cite{ortin}. But the Euler derivative we defined in (\ref{euler}) is not $G^{\m\n}$ but $\sqrt{-g}G^{\m\n}$. So we take our gauge identity as,
\begin{eqnarray}
\Lambda_{\a}\equiv2\nabla_{\b}L^{\b}_{\a}=0.
\label{iden}
\end{eqnarray}
The extra factor 2 is introduced for later convenience. In order to write the above eq. (\ref{iden}) in a more convenient way we note that, the definition of $\Gamma$ (\ref{Gamma}) can be used to write the divergence of Einstein tensor 
\begin{eqnarray}
\nabla_{\m}G^{\m}_{\n}=\p_{\m}G^{\m}_{\n}+\Gamma^{\m}_{\m\a}G^{\a}_{\n}-\Gamma^{\a}_{\m\n}G^{\m}_{\a}
\end{eqnarray}
in the following form
\begin{eqnarray}
\nabla_{\m}G^{\m}_{\n}=(\p_{\m}G^{\m}_{\n}+\frac{1}{2}g^{\m\b}\p_{\a}g_{\m\b}G^{\a}_{\n}-\frac{1}{2}G^{\m\b}\p_{\n}g_{\b\m}).
\label{nabG}
\end{eqnarray}
Now using (\ref{metric}) and (\ref{nabG}) we write the gauge identity (\ref{iden}) in the form
\begin{eqnarray}
\Lambda_{\n}=2\nabla_{\m}L^{\m}_{\n}=
2\nabla_{\m}\sqrt{-g}G^{\m}_{\n}&=&2\sqrt{-g}(\p_{\m}G^{\m}_{\n}+\frac{1}{2}g^{\m\b}\p_{\a}g_{\m\b}G^{\a}_{\n}-\frac{1}{2}G^{\m\b}\p_{\n}g_{\b\m})\nonumber\\
&=&2\p_{\m}\sqrt{-g}G^{\m}_{\n}-\p_{\n}g_{\a\b}\sqrt{-g}G^{\a\b}\nonumber\\
&=&2\p_{\m}L^{\m}_{\n}-\p_{\n}g_{\a\b}L^{\a\b}
\label{ident}
\end{eqnarray}
where we have used the important relation
\begin{eqnarray}
\p_{\m}g=gg^{\a\b}\p_{\m}g_{\a\b}.
\end{eqnarray}
In the metric formulation of gravity the analogy of (\ref{lam}) is expressed as,
\begin{eqnarray}
\Lambda_{\a}(z)=\sum_{s=0}^{n}\int \textrm{d}^3 {\bf x}\frac{\p^s}{\p t^s}\left(\r_{\m\n\a(s)}(x,z)L^{\m\n}(x)\right).
\label{47}
\end{eqnarray}
Comparing this equation with the identity (\ref{ident}), we get the following expressions for the non vanishing generators
\begin{eqnarray}
&&\r_{000(0)}=-\p_0g_{00}\d(x-z)
\label{00}\\
&&\r_{000(1)}=2g_{00}\d(x-z)
\label{01}\\
&&\r_{00k(0)}=-\p_kg_{00}\d(x-z)
\label{k0}\\
&&\r_{00k(1)}=2g_{0k}\d(x-z)
\label{k1}\\
&&\r_{0i0(0)}=-\p_0g_{0i}\d(x-z)+\p_i^z\left(g_{00}\d(x-z)\right)\\
&&\r_{0i0(1)}=g_{0i}\d(x-z)\\
&&\r_{0ik(0)}=-\p_kg_{0i}\d(x-z)+\p_i^z\left(g_{0k}\d(x-z)\right)\\
&&\r_{0ik(1)}=g_{ki}\d(x-z)\\
&&\r_{ij0(0)}=-\p_0g_{ij}\d(x-z)+\p_j^z\left(g_{i0}\d(x-z)\right)+\p_i^z\left(g_{j0}\d(x-z)\right)\\
&&\r_{ijk(0)}=-\p_kg_{ij}\d(x-z)+\p_j^z\left(g_{ik}\d(x-z)\right)+\p_i^z\left(g_{jk}\d(x-z)\right).
\end{eqnarray}
After getting the expressions for the generators it is now straightforward to calculate the diffeomorphism transformation of the metric $g_{\m\n}$ from the general equation (\ref{a}). We write eq. (\ref{a}) for this formulation of gravity as
\begin{eqnarray}
\d g_{\m\n}(x)=\sum_{s=0}^{n}(-1)^s\int {\textrm d}^3 {\bf z} \ \frac{\p^s\varepsilon^{\a}(z)}{\p t^s}\r_{\m\n\a (s)}(x,z).
\label{58}
\end{eqnarray}
From the above equation we write $\d g_{00}$ as,
\begin{eqnarray}
\d g_{00}(x)&=&\int {\textrm d}^3 {\bf z} \ [\varepsilon^0(z)\r_{000(0)}(x,z)+\varepsilon^k(z)\r_{00k(0)}(x,z)\nonumber\\&&-\frac{\p \varepsilon^0}{\p t}(z)\r_{000(1)}(x,z)-\frac{\p \varepsilon^k}{\p t}(z)\r_{000(1)}(x,z)].
\end{eqnarray}
Using the generators (\ref{00},\ref{01},\ref{k0},\ref{k1}) in the above equation we get
\begin{eqnarray}
\d g_{00}&=&-\p_{0}g_{00}\varepsilon^{0}-2g_{00}\p_{0}\varepsilon^{0}-2g_{0k}\p_{0}\varepsilon^{k}-\p_{k}g_{00}\varepsilon^{k}.
\end{eqnarray}
Similarly all other components of the metric $g_{\m\n}$ can be calculated. Combining everything we write the variation as,
\begin{eqnarray}
\d g_{\m\n}=-\p_{\a}g_{\m\n}\varepsilon^{\a}-g_{\m\a}\p_{\n}\varepsilon^{\a}-g_{\a\n}\p_{\m}\varepsilon^{\a}
\label{metric9}
\end{eqnarray}
Above result expresses the gauge transformation of the metric field $g_{\m\n}$. The variation of the inverse metric $g^{\m\n}$ is obtained easily from the above equation (\ref{metric9}) by observing that
\begin{eqnarray}
\d g^{\m\n}=-g^{\m\a}g^{\n\b}\d g_{\a\b}.
\end{eqnarray}
The expression of $\d g^{\m\n}$ we thus find, is written as,
\begin{eqnarray}
\d g^{\m\n}=-\p_{\a}g^{\m\n}\varepsilon^{\a}+g^{\m\a}\p_{\a}\varepsilon^{\n}+g^{\a\n}\p_{\a}\varepsilon^{\m}.
\label{inv}
\end{eqnarray}
We next calculate the diffeomorphism transformation of the connection from its definition (\ref{Gamma}). Making use of (\ref{metric9}) and (\ref{inv}) we find it to be
\begin{eqnarray}
\d\Gamma^{\r}_{\m\n}=-\varepsilon^{\a}\p_{\a}\Gamma^{\r}_{\m\n}+\Gamma^{\a}_{\m\n}\p_{a}\varepsilon^{\r}-\Gamma^{\r}_{\m\a}\p_{\n}\varepsilon^{\a}-\Gamma^{\r}_{\a\n}\p_{\m}\varepsilon^{\a}-\p_{\m}\p_{\n}\varepsilon^{\r}.
\label{inv56}
\end{eqnarray}
The gauge variations of the metric field and the connection are also derived in Appendix (eqs. (\ref{g.t.1}) and (\ref{connc2})) from a point of view of infinitesimal general coordinate transformation. The results are naturally identical with eqs. (\ref{metric9}) and (\ref{inv56}).

\section{Palatini Formulation}
Similar to the first order formulation of the non-Abelian gauge theory there is also a first order version of the Einstein-Hilbert action\cite{mann}
\begin{eqnarray}
S=\int d^4x\mathcal{L}(g,\Gamma)=\int d^4x\sqrt{-g}g^{\m\n}R_{\m\n}(\Gamma).
\end{eqnarray}
where $g$, $\Gamma$ are now treated as independent field variables.
The definition of $R_{\m\n}$ in terms of $\Gamma$ is same as (\ref{ricci}). In our present analysis we take $g_{\m\n}$ and $\Gamma^{\r}_{\m\n}$ as symmetric in $\m$ and $\n$. Variation of the above action with respect to $g_{\m\n}$ and $\Gamma^{\r}_{\m\n}$ gives
\begin{eqnarray}
\d S=\int L^{\m\n}\d g_{\m\n}+\int E^{\m\n}_{\s}\d \Gamma^{\s}_{\m\n}.
\end{eqnarray}
The expressions of the Euler derivatives $L^{\m\n}$ and $E^{\m\n}_{\s}$ are given below
\begin{eqnarray}
L^{\m\n}&=&\sqrt{-g}(R^{\m\n}-\frac{1}{2}g^{\m\n}R)
\label{66}\\
E^{\m\n}_{\s}&=&(\sqrt{-g}g^{\m\n})_{,\s}+\sqrt{-g}g^{\r\n}\Gamma^{\m}_{\r\s}+\sqrt{-g}g^{\m\r}\Gamma^{\n}_{\r\s}-\sqrt{-g}g^{\m\n}\Gamma^{\r}_{\r\s}\nonumber\\
&&-\frac{1}{2}[(\sqrt{-g}g^{\m\b})_{,\b}+\sqrt{-g}g^{\r\b}\Gamma^{\m}_{\r\b}]\d^{\n}_{\s}-\frac{1}{2}[(\sqrt{-g}g^{\n\b})_{,\b}+\sqrt{-g}g^{\r\b}\Gamma^{\n}_{\r\b}]\d^{\m}_{\s}\nonumber\\
\label{67}
\end{eqnarray}
leading to the equations of motion,
\begin{eqnarray}
&&L^{\m\n}=0\\
&&E^{\m\n}_{\s}=0.
\end{eqnarray}
The metric compatibility condition (\ref{metric}) can be derived from the second equation of motion\cite{ortin} except for two dimension. The issues related to two dimension may be found in \cite{mann}.
After obtaining the Euler derivatives we now give the gauge identity for the Palatini formulation
\begin{eqnarray}
\Lambda_{\a}&\equiv&-L^{\m\n}\p_{\a}g_{\m\n}+2\p_{\n}(g^{\n\s}L_{\a\s})\nonumber\\
&&-E^{\m\n}_{\r}\p_{a}\Gamma^{\r}_{\m\n}-\p_{\r}(E^{\m\n}_{\a}\Gamma^{\r}_{\m\n})+2\p_{\m}(E^{\m\n}_{\r}\Gamma^{\r}_{\a\n})\p_{\m}\p_{\n}E^{\m\n}_{\a}=0.
\label{giden}
\end{eqnarray}
Note that in the metric formulation $E^{\m\n}_{\s}$ is identically zero and eq. (\ref{giden}) reduces to the identity (\ref{ident}) which came from the double contraction of the Bianchi identity (\ref{bianchi}). It is worthwhile to mention that there is also a Bianchi identity\cite{ortin} valid for Palatini formulation
\begin{eqnarray}
\nabla_{\eta}R^{\lambda}_{ \ \m\n\kappa}+\nabla_{\n}R^{\lambda}_{ \ \m\kappa\eta}+\nabla_{\kappa}R^{\lambda}_{ \ \m\eta\n}=0.
\end{eqnarray}
 The relation (\ref{47}) is now rewritten to include the independent Euler derivatives as,
\begin{eqnarray}
\Lambda_{\a}(z)=\sum_{s=0}^{n}\int \textrm{d}^3 {\bf x}\frac{\p^s}{\p t^s}\left(\r_{\m\n\a(s)}(x,z)L^{\m\n}(x)+\r^{\s}_{\m\n\a(s)}(x,z)E_{\s}^{\m\n}(x)\right).
\label{69}
\end{eqnarray}
Using the explicit expressions for the Euler derivatives (\ref{66},\ref{67}), the generators are read off by comparing (\ref{giden}) with (\ref{69}), expectedly, the generators $\r_{\m\n\a(s)}$ are identical to the generators of the metric formulation, exactly as happened for the gauge theory. The expressions of the other generators are given below
\begin{eqnarray}
&&\r^{0}_{000(0)}=-\p_0\Gamma^0_{00}\d(x-z)-\p_m^z\left(\Gamma^m_{00}\d(x-z)\right)
\label{n1}\\
&&\r^{0}_{00k(0)}=-\p_k\Gamma^0_{00}\d(x-z)
\label{n2}\\
&&\r^{0}_{000(1)}=\Gamma^0_{00}\d(x-z)
\label{n3}\\
&&\r^{0}_{00k(1)}=2\Gamma^0_{0k}\d(x-z)
\label{n4}\\
&&\r^{0}_{000(2)}=-\d(x-z)
\label{n5}\\
&&\r^{0}_{0i0(0)}=-\p_0\Gamma^0_{0i}\d(x-z)-\p_m^z\left(\Gamma^m_{0i}\d(x-z)\right)+\p_i^z\left(\Gamma^0_{00}\d(x-z)\right)\\
&&\r^{0}_{0ik(0)}=-\p_k\Gamma^0_{0i}\d(x-z)+\p_i^z\left(\Gamma^0_{0k}\d(x-z)\right)\\
&&\r^{0}_{0ik(1)}=\Gamma^0_{ki}\d(x-z)\\
&&\r^{0}_{0i0(1)}=-\p_i^z\d(x-z)\\
&&\r^{0}_{ij0(0)}=-\p_0\Gamma^0_{ij}\d(x-z)+\p_j^z\left(\Gamma^0_{i0}\d(x-z)\right)+\p_i^z\left(\Gamma^0_{0j}\d(x-z)\right)\nonumber\\&&\quad\quad\quad\quad-\p_m^z\left(\Gamma^m_{ij}\d(x-z)\right)-\p_i^z\p_j^z\d(x-z)\\
&&\r^{0}_{ijk(0)}=-\p_k\Gamma^0_{ij}\d(x-z)+\p_j^z\left(\Gamma^0_{ik}\d(x-z)\right)+\p_i^z\left(\Gamma^0_{kj}\d(x-z)\right)\\
&&\r^{0}_{ij0(1)}=-\Gamma^0_{ij}\d(x-z)\\
&&\r^{k}_{000(0)}=-\p_0\Gamma^k_{00}\d(x-z)\\
&&\r^{k}_{00m(0)}=-\p_m\Gamma^k_{00}\d(x-z)-\p_p^z\left(\Gamma^p_{00}\d(x-z)\right)\d^k_m\\
&&\r^{k}_{00m(1)}=-\Gamma^0_{00}\d^k_m\d(x-z)+2\Gamma^k_{m0}\d(x-z)\\
&&\r^{k}_{000(1)}=2\Gamma^k_{00}\d(x-z)\\
&&\r^{k}_{00m(2)}=-\d(x-z)\\
&&\r^{k}_{0i0(0)}=-\p_0\Gamma^k_{0i}\d(x-z)+\p_i^z\left(\Gamma^k_{00}\d(x-z)\right)\\
&&\r^{k}_{0im(0)}=-\p_m\Gamma^k_{0i}\d(x-z)-\d^k_m\p_p^z\left(\Gamma^p_{0i}\d(x-z)\right)+\p_i^z\left(\Gamma^k_{0m}\d(x-z)\right)\nonumber\\
\\
&&\r^{k}_{0im(1)}=-\d^k_m\Gamma^0_{0i}\d(x-z)+\Gamma^k_{mi}\d(x-z)-\p_i^z\d(x-z)\d^k_m\\
&&\r^{k}_{0i0(1)}=\Gamma^k_{0i}\d(x-z)\\
&&\r^{k}_{ij0(0)}=-\p_0\Gamma^k_{ij}\d(x-z)+\p_i^z\left(\Gamma^k_{0j}\d(x-z)\right)+\p_j^z\left(\Gamma^k_{0i}\d(x-z)\right)\\
&&\r^{k}_{ijm(0)}=-\p_m\Gamma^k_{ij}\d(x-z)+\p_i^z\left(\Gamma^k_{jm}\d(x-z)\right)+\p_j^z\left(\Gamma^k_{im}\d(x-z)\right)\nonumber\\&&\quad\quad\quad\quad-\d^k_m\p_p^z\left(\Gamma^p_{ij}\d(x-z)\right)-\d^k_m\p_i^z\p_j^z\d(x-z)\\
&&\r^{k}_{ijm(1)}=-\d^k_m\Gamma^0_{ij}\d(x-z).
\end{eqnarray}
These generators are now used to calculate the diffeomorphism transformation of the fields $g_{\m\n}$ and $\Gamma^{\r}_{\m\n}$ from the following equations
\begin{eqnarray}
\d g_{\m\n}(x)=\sum_{s=0}^{n}(-1)^s\int {\textrm d}^3 {\bf z}\frac{\p^s\varepsilon^{\a}}{\p t^s}(z)\r_{\m\n\a (s)}(x,z)\\
\d \Gamma_{\m\n}^{\r}(x)=\sum_{s=0}^{n}(-1)^s\int {\textrm d}^3 {\bf z}\frac{\p^s\varepsilon^{\a}}{\p t^s}(z)\r_{\m\n\a (s)}^{\r}(x,z).
\label{connc}
\end{eqnarray}
Since the generators $\r_{\m\n\a (s)}$ are same as the generators of the metric formulation of gravity, the gauge variation of the metric is same as (\ref{metric9}). From eq. (\ref{connc}) we write $\d \Gamma^0_{00}$ explicitly
\begin{eqnarray}
\d \Gamma^0_{00}(x)&=&\int {\textrm d}^3 {\bf z} \ [\varepsilon^0(z)\r^0_{000(0)}(x,z)+\varepsilon^k(z)\r^0_{00k(0)}(x,z)\nonumber\\&&-\frac{\p\varepsilon^0}{\p t}(z)\r^0_{000(1)}(x,z)-\frac{\p\varepsilon^k}{\p t}(z)\r^0_{00k(1)}(x,z)+\frac{\p^2\varepsilon^0}{\p t^2}(z)\r^0_{000(2)}(x,z)]\nonumber\\
\end{eqnarray}
using (\ref{n1},\ref{n2},\ref{n3},\ref{n4},\ref{n5}) in the above equation we get
\begin{eqnarray}
\d \Gamma^0_{00}=-\varepsilon^{\a}\p_{\a}\Gamma^0_{00}+\Gamma^{\a}_{00}\p_{\a}\varepsilon^0-2\Gamma^{0}_{0\a}\p_0\varepsilon^{\a}-\p_0^2\varepsilon^0.
\end{eqnarray}
All other components of the connection can be calculated in a similar manner. The results thus obtained are written in the following way
\begin{eqnarray}
\d\Gamma^{\r}_{\m\n}=-\varepsilon^{\a}\p_{\a}\Gamma^{\r}_{\m\n}+\Gamma^{\a}_{\m\n}\p_{a}\varepsilon^{\r}-\Gamma^{\r}_{\m\a}\p_{\n}\varepsilon^{\a}-\Gamma^{\r}_{\a\n}\p_{\m}\varepsilon^{\a}-\p_{\m}\p_{\n}\varepsilon^{\r}.
\end{eqnarray}
Thus in the Palatini formulation, the gauge transformation of the connection is derived independently from the gauge variation of the metric. The same result is derived in Appendix (eq. (\ref{connc2})) using the transformation of the metric (\ref{metric9}).
\section{Conclusions}
We have studied the diffeomorphism symmetries of the general relativity theory for both the second order (metric) and the first order (Palatini) formulations of gravity. Some identities were obtained from which the generators of the transformations were found. The diffeomorphism transformations of the metric and the connection for both approaches were systematically derived. The results thus obtained were compatible with each other. It is noteworthy that the A-D-M splitting, which is essential for discussing diffeomorphism symmetries, is bypassed.

 It is worthwhile to pursue the consequence of symmetry analysis for other approaches of gravity since it is a well known fact that Einstein-Hilbert action is not the unique action which is invariant under general coordinate transformation. Even for a more general Lovelock gravity formulation\cite{lov} it is shown that classical results obtained from metric and Palatini formulations are completely equivalent\cite{jabbari}. In presence of different other terms in an action (higher order gravity theory\cite{carrollS,vollick,nojiri}) apart from the standard Einstein Hilbert term, metric and Palatini formulations are not equivalent in general\cite{vollicka,flanagan}. For such an action the present symmetry studies may give new insights about the problem.
\appendix
\section{Appendix}
Here we give a brief derivation of the diffeomorphism transformation of the fields. Under a general coordinate transformation, $g_{\m\n}$ transforms as a covariant second rank tensor i. e.
\begin{eqnarray}
g'_{\m\n}(x')=\frac{\p x^{\a}}{\p x'^{\m}}\frac{\p x^{\b}}{\p x'^{\n}}g_{\a\b}(x)
\label{trag}
\end{eqnarray}
Now we consider an infinitesimal coordinate transformation
\begin{eqnarray}
x^{\m}\rightarrow x'^{\m}=x^{\m}+\varepsilon^{\m}(x)
\label{eps}
\end{eqnarray}
under which, we write from (\ref{trag})
\begin{eqnarray}
g'_{\m\n}(x')=g_{\m\n}(x)-g_{\m\a}(x)\p_{\n}\varepsilon^{\a}-g_{\a\n}(x)\p_{\m}\varepsilon^{\a}+\mathcal{O}(\varepsilon^2)
\label{g1}
\end{eqnarray}
A Taylor expansion of the r. h. s. of the above equation, using (\ref{eps}), gives
\begin{eqnarray}
g'_{\m\n}(x')=g'_{\m\n}(x+\varepsilon)=g'_{\m\n}(x)+\p_{\a}g_{\m\n}\varepsilon^{\a}+\mathcal{O}(\varepsilon^2)
\label{g2}
\end{eqnarray}
Combining (\ref{g1}) and (\ref{g2}) we get
\begin{eqnarray}
\d g_{\m\n}(x)=g'_{\m\n}(x)-g_{\m\n}(x)=-\p_{\a}g_{\m\n}\varepsilon^{\a}-g_{\m\a}\p_{\n}\varepsilon^{\a}-g_{\a\n}\p_{\m}\varepsilon^{\a}
\label{g.t.1}
\end{eqnarray}
This can be written in a covariant notation also\cite{Landau}
\begin{eqnarray}
\d g_{\m\n}=-\nabla_{\m}\varepsilon_{\n}-\nabla_{\n}\varepsilon_{\m}
\end{eqnarray}
where the definition of connection (\ref{Gamma}) has been used. In a similar way the variation of the inverse metric $g^{\m\n}$ is also obtained
\begin{eqnarray}
\d g^{\m\n}=-\p_{\a}g^{\m\n}\varepsilon^{\a}+g^{\m\a}\p_{\a}\varepsilon^{\n}+g^{\a\n}\p_{\a}\varepsilon^{\m}
\label{g.t.2}
\end{eqnarray}
Using (\ref{g.t.1}) and (\ref{g.t.2}) it is now straightforward to calculate the diffeomorphism transformation of the connection from (\ref{Gamma}). We find it to be
\begin{eqnarray}
\d\Gamma^{\r}_{\m\n}=-\varepsilon^{\a}\p_{\a}\Gamma^{\r}_{\m\n}+\Gamma^{\a}_{\m\n}\p_{a}\varepsilon^{\r}-\Gamma^{\r}_{\m\a}\p_{\n}\varepsilon^{\a}-\Gamma^{\r}_{\a\n}\p_{\m}\varepsilon^{\a}-\p_{\m}\p_{\n}\varepsilon^{\r}
\label{connc2}.
\end{eqnarray}
This equation together with (\ref{g.t.1}) give the gauge variation of the two most important quantities of general theory of relativity.
\section*{Acknowledgment}
We thank Dr. R. Banerjee for suggesting this investigation and also for discussions.

\end{document}